\documentclass{article}

\usepackage{color,array,amsthm}
\usepackage{graphicx}

\usepackage{amsmath}
\usepackage{graphicx}
\usepackage{algpseudocode}
\usepackage{algorithm}
\usepackage{amsfonts}
\usepackage{longtable}
\usepackage{array}        
\usepackage{booktabs}     
\usepackage{tabularx}     
\usepackage{multicol}
\usepackage{multirow}

\usepackage{balance}  

\usepackage{arxiv}
\usepackage[utf8]{inputenc} 
\usepackage[T1]{fontenc}    
\usepackage{hyperref}       
\usepackage{url}            
\usepackage{nicefrac}       
\usepackage{microtype}      
\usepackage{lipsum}		
\usepackage{graphicx}
\usepackage{doi}

\MakeRobust{\Call}

\newcommand{\myspace}{\vspace{1.5mm}}

\newcommand\orcid[1]{\href{https://orcid.org/#1}{\includegraphics[scale=0.09]{ORCIDiD_icon128x128.png}}}

\author{ \href{https://orcid.org/0009-0000-1598-8056}{\includegraphics[scale=0.06]{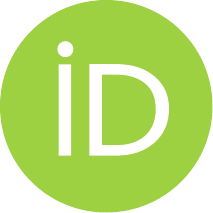}\hspace{1mm}Blaž Pšeničnik}\\
	Computer Architecture and Languages Laboratory\\
	Faculty of Electrical Engineering and Computer Science\\
	University of Maribor\\
	\texttt{blaz.psenicnik1@um.si} \\
	\And
	\href{https://orcid.org/0009-0006-2116-4233}{\includegraphics[scale=0.06]{orcid.pdf}\hspace{1mm}Rene Mlinarič} \\
	Computer Architecture and Languages Laboratory\\
	Faculty of Electrical Engineering and Computer Science\\
	University of Maribor\\
	\texttt{rene.mlinaric@student.um.si} \\
 	\And
	\href{https://orcid.org/0000-0001-5864-3533}{\includegraphics[scale=0.06]{orcid.pdf}\hspace{1mm}Janez Brest} \\
	Computer Architecture and Languages Laboratory\\
	Faculty of Electrical Engineering and Computer Science\\
	University of Maribor\\
	\texttt{janez.brest@um.si} \\
  	\And
 	\href{https://orcid.org/0000-0002-7595-2845}{\includegraphics[scale=0.06]{orcid.pdf}\hspace{1mm}Borko Bošković} \\
	Computer Architecture and Languages Laboratory\\
	Faculty of Electrical Engineering and Computer Science\\
	University of Maribor\\
	\texttt{borko.boskovic@um.si} \\
}

\begin{document}
\title{Dual-Step Optimization for Binary Sequences with High Merit Factors} 
\maketitle


\begin{abstract}The problem of finding aperiodic low auto-correlation binary sequences (LABS) presents a significant computational challenge, particularly as the sequence length increases. Such sequences have important applications in communication engineering, physics, chemistry, and cryptography. This paper introduces a novel dual-step algorithm for long binary sequences with high merit factors. The first step employs a parallel algorithm utilizing skew-symmetry and restriction classes to generate sequence candidates with merit factors above a predefined threshold. The second step uses a priority queue algorithm to refine these candidates further, searching the entire search space unrestrictedly. By combining GPU-based parallel computing and dual-step optimization, our approach has successfully identified new best-known binary sequences for all lengths ranging from 450 to 527, with the exception of length 518, where the previous best-known value was matched with a different sequence. This hybrid method significantly outperforms traditional exhaustive and stochastic search methods, offering an efficient solution for finding long sequences with good merit factors.
\end{abstract}


\section{INTRODUCTION}
The aperiodic low auto-correlation binary sequences (LABS) problem presents a formidable computational challenge. Golay~\cite{golay72} initially presented the problem in 1972. Before Golay's formulation, Littlewood~\cite{littlewood66} explored the norms of polynomials with $\pm1$ coefficients on the unit circle of the complex plane, a problem that is similar in nature to our problem. Binary sequences with high merit factors, and therefore low auto-correlation properties, have important applications in digital communications~\cite{jedwab2004survey,katz2024moments} (e.g., they
allow the efficient separation of signals from noise), as well as in 
physics~\cite{Lib-OPUS-labs-1987-JourPhys-Bernasconi}, 
chemistry, 
cryptography, etc. 

A binary sequence of length $L$ is defined as $S(L)=\{s_1, s_2, ..., s_L\}$, where $s_i \in \{+1, -1\}$. The auto-correlation function is then equal to $C_k(S)=\sum_{i=1}^{L-k} s_i s_{i+k}$ while the energy can be calculated as $E(S)=\sum_{k=1}^{L-1} C_k^2(S)$. The objective of the problem is to maximize the merit factor $F$~\cite{Mullen13HANDBOOKMERITFACTOR}:
\begin{equation}\label{eq:merit_factor}
F(S)=\frac{L^2}{2E(S)}.
\end{equation}

For smaller instances of the problem, a complete search can be used to find optimal solutions, but this approach quickly becomes infeasible, so stochastic methods are used for longer sequences, which produce good solutions. Due to the exponential growth of the problem's search space, which is $2^L$, reducing it can be highly beneficial. A skew-symmetric sequence~\cite{golay72}, of odd-length $L=2k+1$ must satisfy:
\begin{equation}\label{eq:skew_symmetry}
s_{(k+1)+i}=(-1)^i s_{(k+1)-i},\quad i = 1,2,\,\cdots,k.
\end{equation}

This restriction effectively reduces the search space to approximately $2^{(L/2)}$ and guarantees that all $C_k(S)=0$, for all odd values of $k$, which lowers the overall value of $E(S)$ while increasing the merit factor. It must be noted that a skew-symmetric solution might not be optimal for a given instance of the problem.

Furthermore, we can reduce search space to approximately $2^{(L/2-p)}$ by fixating first $p$ elements in the sequence. The first $p$ elements are set using partitions (restriction classes)~\cite{dimitrov22} of length $p$ with minimal or normalized potentials, the next $k - p + 1$ elements are free to change, while the last $k$ elements are determined by the skew-symmetry rule, as shown in~\eqref{eq:partitions}. This effectively sets some elements in the auto-correlation function to small values, which lowers the overall energy.

\begin{align}\label{eq:partitions}
S(L)=\underbrace{s_1s_2 \cdots s_{p}}_{p} \underbrace{s_{p+1}s_{p+2} \cdots s_{k-1}s_{k}s_{k+1}}_{k-p+1} \underbrace{s_{k+2}s_{k+3} \cdots s_{L-1}s_{L}}_{k} 
\end{align}


In this paper, we propose a dual-step algorithm for searching long binary sequences with high merit factors. The first step consists of a parallel algorithm, that utilizes skew-symmetry and restriction classes. This step produces candidates with a merit factor higher than a predefined limit. The second step consists of a priority queue algorithm, that further improves the merit factors of candidates from the first step. It must be noted that construction methods~\cite{borwein04, Jedwab13} can swiftly produce sequences of a certain quality.  We introduced these sequences as candidates to the second step (due to them being non-skew-symmetric) in the hopes of increasing the merit factor even further.

According to the above, the main contributions of this paper are:
\begin{itemize}
    \item An algorithm that produces sequences with good merit factors, which consists of two steps:
    \begin{itemize}
        \item A parallel algorithm for searching binary sequences with good merit factors that takes advantage of GPGPU devices, based on the work of M. Dimitrov~\cite{dimitrov22}.
        \item An algorithm using a priority queue, that searches the whole search space without any restrictions.
    \end{itemize}
    \item New best-known binary sequences of lengths $450 \le L \le 527$, with the exception of $518$, where the previous best-known value was matched with a different sequence.
\end{itemize}

The remainder of this paper is organized as follows. The related work is described in Section \nolinebreak~\ref{sec:related_work}. The proposed algorithm and both steps are described in Section~\ref{sec:algo}. The description of experiments and results are described in Section~\ref{sec:results}. Finally, the conclusion is written in Section~\ref{sec:conclusions}.

\section{RELATED WORK}\label{sec:related_work}
In literature, binary sequences are characterized by two types of auto-correlation functions: periodic and aperiodic~\cite{Schmidt2016}. Aperiodic auto-correlation is often considered a more realistic reflection of sequence behavior in practical systems. Research on LABS searches explores two distinct approaches, each optimizing a different metric: one prioritizes minimizing the peak sidelobe level (PSL)~\cite{Mow2015, Chen2022, Kimura2024}, while the other focuses on maximizing the merit factor ($F$). In this paper, we concentrate on finding long binary sequences that exhibit a high merit factor.

\begin{figure}
\centering
\includegraphics[width=0.6\linewidth]{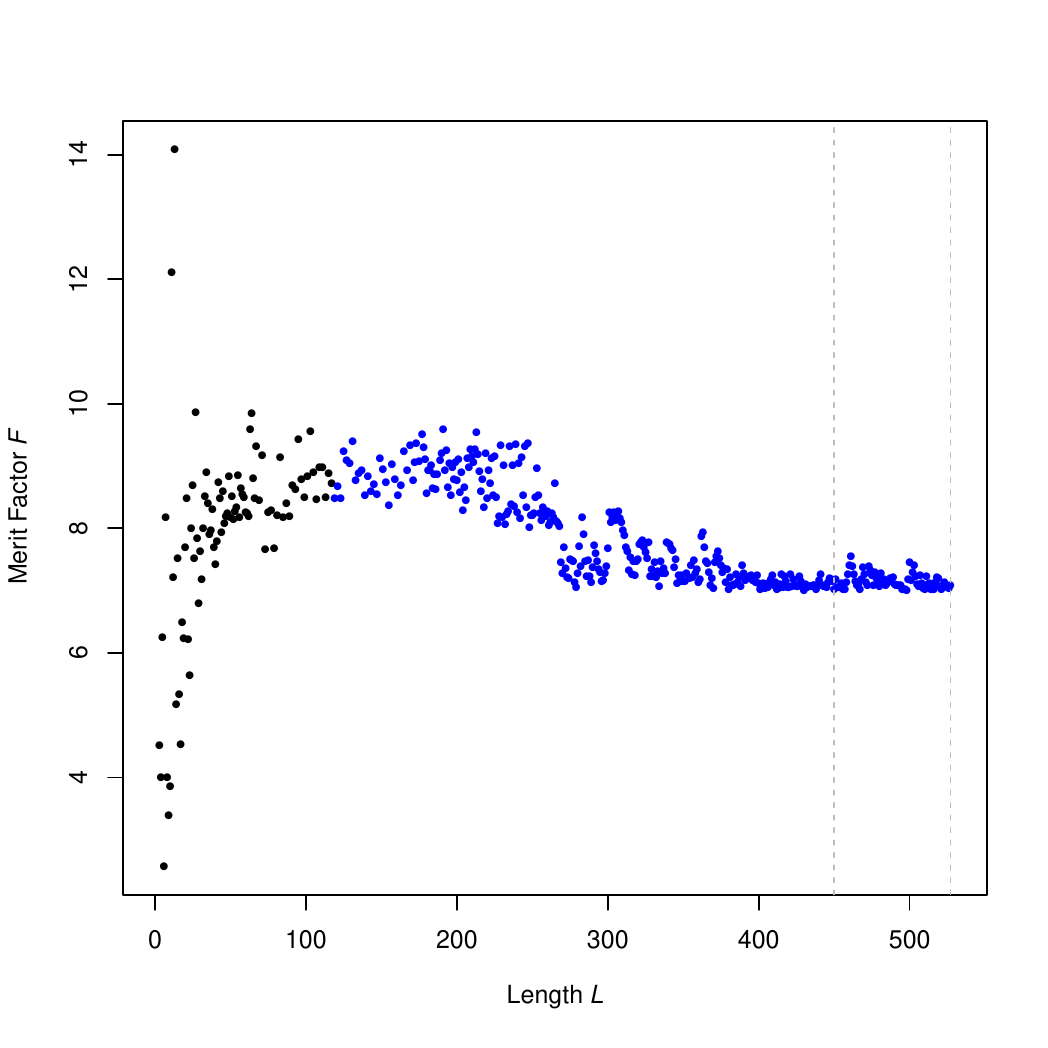}
\caption{\label{fig:oldmf}Largest known merit factors ($F$) for $L \le 527$. Black symbols are optimal solutions from exhaustive searches~\cite{Packebusch16}, blue symbols are solutions from stochastic approaches~\cite{Boskovic17,Boskovic24, dimitrov22}. The highlighted range $450 \le L \le 527$ indicates where the focus of this work is.}
\end{figure}

Three approaches exist to find such (sub)-optimal sequences. The first is the construction method. With the help of this approach, it is possible to construct non-optimal sequences with a certain quality in a very short time. As shown in ~\cite{Baden11,Jedwab13}, this approach is suitable for longer sequences with a merit factor value of around $6.3421$ or $6.4382$. The second is an exact search that allows finding the optimal sequences. This approach is time-consuming and inappropriate for long sequences. Currently, it provides the optimal sequences for $L\le 66$ and the skew-symmetric optimal sequences for $L \le 119$~\cite{Packebusch16}. The quantum approximate optimization algorithm (QAOA) is proposed in~\cite{shaydulin2024evidence} to tackle LABS for $L \le 40$ and the observed runtime of QAOA with fixed parameters scales better than state-of-the-art exact solvers. The third is a stochastic search that can find good sequences that are not necessarily optimal. From this, it is evident that the construction method and a stochastic approach are more suitable for longer sequences. Stochastic algorithms are helpful for sequences where we can achieve better merit factors than those achieved by construction methods.

The following stochastic approaches were used in literature: local search~\cite{Farnane18,dimitrov21}, tabu search~\cite{Halim08}, evolutionary algorithm combined with tabu search~\cite{Gallardo09}, self-avoiding walk technique~\cite{Boskovic17,Boskovic24}, guiding the search process with a priority queue~\cite{Brest18,Brest22}, etc. The local search algorithms in~\cite{Boskovic17, dimitrov22} found sequences for $L \le 268$ with merit factors greater than 8 and for $L \le 527$ greater than 7. Since exhaustive search is inapplicable to large values of $L$, the authors introduced restriction classes in~\cite{dimitrov22}. This modification also enabled the parallelization of the search into multiple non-overlapping instances. The authors also showed that algorithms for skew-symmetric binary sequences of odd length can be transformed to search pseudo-skew-symmetric binary sequences of even lengths.

Analyzing the stochastic algorithm on short sequences makes establishing a prediction model of stopping conditions possible. With its help and parallel computing on the graphics processing units~\cite{Boskovic24}, the stochastic solver found optimal skew-symmetric sequences with an estimated probability of 99\% for $L \le 223$ and skew-symmetric sequences with merit factors greater than 9 for $L ~\le 247$. Fig.~\ref{fig:oldmf} shows the largest known merit factors to date, with solutions from exhaustive searches~\cite{Packebusch16} and stochastic approaches~\cite{Boskovic17,Boskovic24, dimitrov22}.

\begin{figure*}[tbh]
\centering
\includegraphics[width=0.9\linewidth]{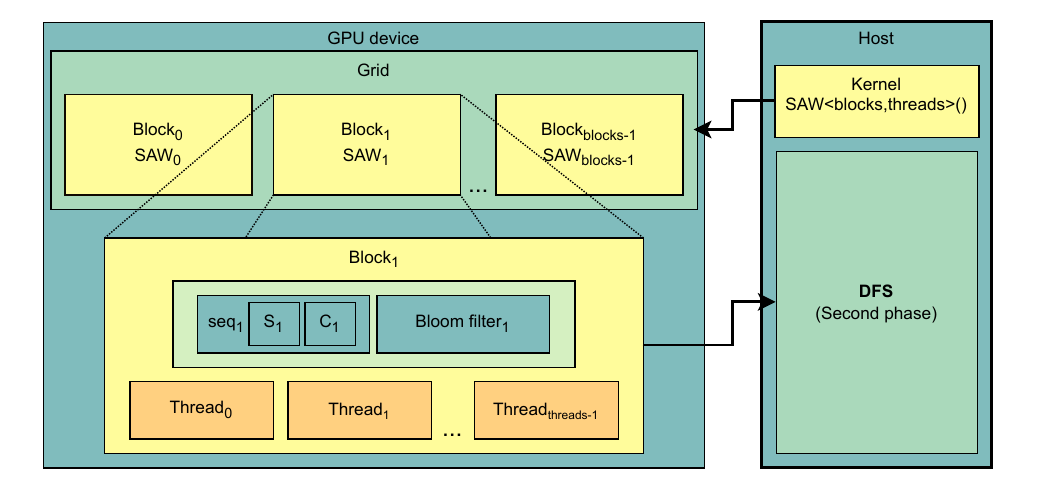} 
\caption{\label{fig:gpgpu}The arhitecture of SAW algorithm.}
\end{figure*}

In contrast to related works, we designed a hybrid approach that uses two steps. We adopted a combination of established techniques. The first step uses parallel self-avoiding walks on GPU devices, skew-symmetry, and restriction classes to reach sequences with good merit factors, based on the work in~\cite{dimitrov22,Boskovic24}. We used these solutions as initial sequences in the second step. The search process, guided by a priority queue, similar to the works of J. Brest et al.~\cite{Brest18,Brest22}, is used in this step to reach sequences with even better merit factors, exploring the whole search space and ignoring skew-symmetry and partitions. In this work, we will focus on sequences of lengths $450 \le L \le 527$.

\section{OUR WORK}\label{sec:algo}
In order to achieve high merit factors for long binary sequences, we propose a dual-step algorithm. The first step modifies the algorithm proposed in~\cite{dimitrov22} to use General-Purpose computing on Graphical Processing Units (GPGPU) for parallel self-avoiding walks. It takes advantage of skew-symmetry and restriction classes. However, the limitations of the Graphical Processing Unit (GPU) memory pose a challenge for this approach. To prevent revisiting previously explored sequences, the algorithm stores their hashes. Consequently, the walk length is directly limited by available memory. To mitigate this issue and increase the search depth, a Bloom Filter~\cite{bloom70} was used.

The first strep starts by running many parallel self-avoiding walks $\{\mathit{SAW_0},\mathit{SAW_1},...,\mathit{SAW_n}\}$. Each walk is independent of the other,  meaning efficient parallelization is possible on GPU devices. Each self-avoiding walk on the device is performed inside a block, as shown in Fig.~\ref{fig:gpgpu}, effectively realizing a breadth-first search over the restricted search space. Each block runs the algorithm described in Algorithm~\ref{algo:saw}. 
\begin{algorithm}[tbh]
\caption{First step for searching skew-symmetric binary sequences with partitions}\label{algo:saw}
\begin{small}
\begin{algorithmic}[1]
    \State $\mathbb{BF} \gets \emptyset$ \Comment{Create an empty Bloom Filter }
    \State $seq \gets \Call{init\_partitioned\_sequence}$
    \myspace
    \State $\mathbb{BF}.\Call{insert}{seq}$
    \myspace
    \State $e \gets \Call{energy}{seq}$
    \myspace
    \State $it \gets 0$
    \myspace
    \While{$it < T_i $}
        \State $it \gets it + 1$
        \myspace
        \State $i, delta \gets \Call{best\_neighbour}{seq}$
        \myspace
        \If {$i = -1$}
            \State \textbf{break}
        \EndIf
        \myspace
        \State $\Call{sequence\_flip\_skew}{seq, i}$
        \State $\mathbb{BF}.\Call{insert}{seq}$
        \myspace
        \State $e \gets e + delta$
        \myspace
        \If {$e < E_l$}
            \State $\Call{DFS}{seq}$ \Comment{Call DFS with the newly found candidate}
        \EndIf
    \EndWhile
\end{algorithmic}
\end{small}
\end{algorithm}

\begin{figure*}[tbh]
\centering
\includegraphics[width=0.99\linewidth]{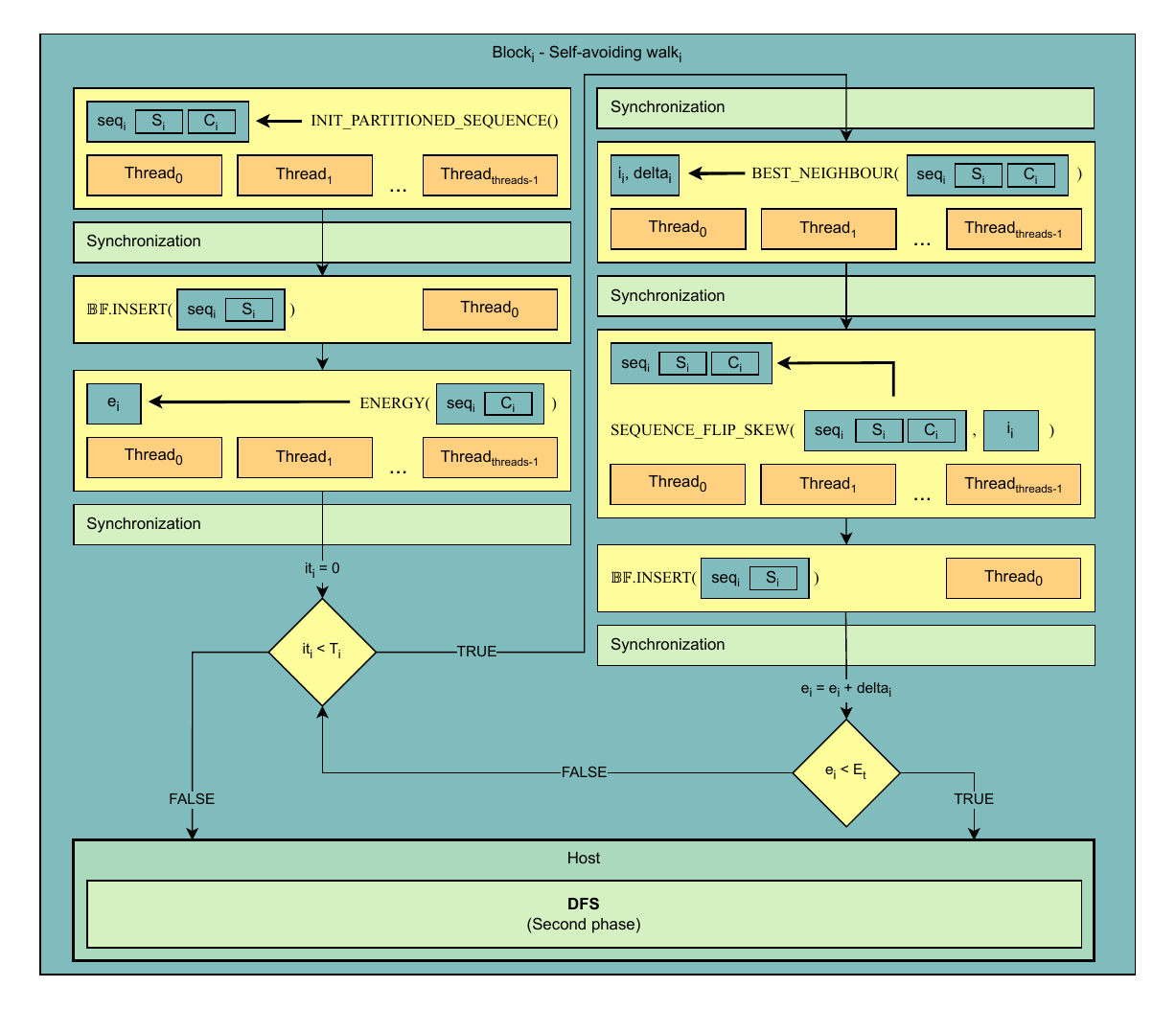} 
\caption{\label{fig:gpgpu_block}Parallel self-avoiding walk inside a block.}
\end{figure*}
\begin{figure}[tbh]
\centering
\includegraphics[width=0.5\linewidth]{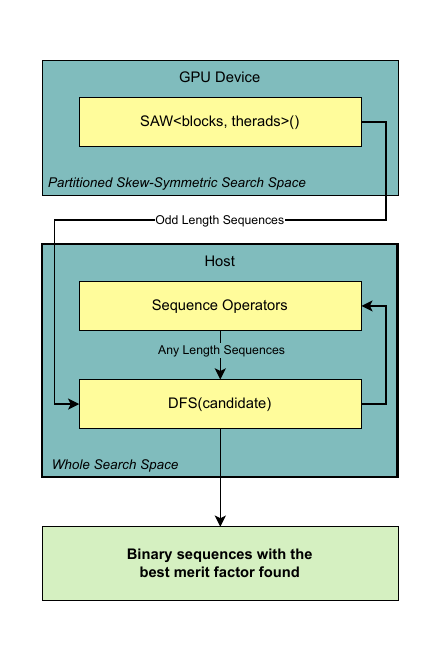}
\caption{\label{fig:flow} Workflow of the two-phase search, that combines the algorithms.} 
\end{figure}

A block consists of multiple threads with shared memory. With these threads, functions in the walk are parallelized and use reduction for efficient computation. The algorithm within one GPU block is depicted in Fig.~\ref{fig:gpgpu_block}. The starting pivot is initialized by the {\small $\mathrm{INIT\_PARTITIONED\_SEQUENCE}$}, which creates a random skew-symmetric sequence belonging to a unique partition (restriction class) of the block as described in~\cite{dimitrov22}. This ensures that each block operates on a unique, non-overlapping subset of the search space, preventing redundancy. As a result, multiple threads or blocks do not explore the same regions, which would otherwise waste computational resources. The length of the walk is limited by $T_i$ due to memory limitations. The value of the parameter was fixed to $8\cdot(L+1)/2$ as this offers best asymptotic average case performance, as shown in~\cite{Boskovic17}. The {\small $\mathrm{BEST\_NEIGHBOUR}$} function performs a parallel neighborhood search using threads and returns the index $i$ of the unexplored neighbor with the lowest energy change $\mathit{delta}$. A neighbor is a binary sequence within the same partition, differing from the pivoting sequence by a single flip, maintaining skew-symmetry and the original partition. The function uses linear time and space complexity to evaluate a neighbor, as described in~\cite{dimitrov21}. After each iteration, the best neighbor becomes the pivot by flipping the element at indexes $i$ and $(k+1)-i$ in the sequence in linear time, as shown in~\cite{dimitrov21}. The proposed algorithm in~\cite{dimitrov22} simultaneously searches for skew-symmetric binary sequences of odd length $L$ and pseudo-skew-symmetric binary sequences with even lengths $L-1$ and $L+1$, which introduces branching. In the design of our algorithm, we intentionally minimized the use of conditional statements to avoid performance degradation due to warp divergence by only searching for skew-symmetric sequences, which allows all threads within a warp to follow a uniform execution path. Searching for sequences of even lengths is handled by using sequence operators~\cite{dimitrov22} and the second step, as shown in Fig.~\ref{fig:flow}, by appending or removing elements on each end of the sequence.

Due to constraints imposed by skew-symmetry and restriction classes, the search might exclude certain better solutions. To mitigate this problem, a second step was introduced, which uses a priority queue to explore the whole search space, as described in Algorithm~\ref{algo:pq}. The idea of using a priority queue to guide the search process was already presented in~\cite{Brest18, Brest22}, but the search space was limited to skew-symmetric sequences only. This method differs from the first step by employing a depth-first search (DFS) over the local search space of a candidate increasing the $F$ value further. Because this step is computationally and memory intensive, only sequences with a low enough energy (i.e., high merit factor) are considered candidates. This is constrained by the $E_l$ parameter in Algorithm~\ref{algo:saw}, which effectively sieves the sequences produced by the first step. In each iteration, the sequence with the lowest energy is dequeued and becomes a pivot. All unexplored neighbors of that sequence are evaluated and inserted back as possible future pivots. The number of iterations is controlled by the $T_u$ parameter, which limits the maximum number of steps since the last energy improvement, effectively implementing dynamic search depth. The function {\small $\mathrm{SEQUENCE\_FLIP}$} is instrumental, offering linear time and space complexity for flipping the $i$-th element of a sequence, as documented in~\cite{dimitrov20}. 
This function differs from the one mentioned above because it doesn't maintain skew-symmetry. The function {\small $\mathrm{MAKE\_ROTATIONS}$} rotates the current pivoting sequence left and right by at most $T_r$ elements and adds the rotated sequences to the $\mathbb{PQ}$. A rotation is considered a shift of all elements while moving the first or last element to the start or end of the sequence, maintaining its length. This was done because rotated sequences have a similar energy $E$ to the original one, which are unreachable by the neighborhood search. It must be noted that sequences returned by the second step are odd-length and not necessarily skew-symmetric or belonging to the original partition of the candidate. Moreover, the algorithm outlined in Algorithm~\ref{algo:pq} can be parallelized across many CPU cores independently to increase performance.

\begin{algorithm}[th]
\caption{Second step using a priority queue}\label{algo:pq}
\begin{small}
\begin{algorithmic}[1]
\Require{$candidate$} \Comment{Candidate from the first phase (SAW)}
\Ensure{$seq_{best}$} \Comment{Binary sequence with the best merit factor found}
\myspace
    \State $\mathbb{PQ}, \mathbb{H} \gets \emptyset, \emptyset\ $ \Comment{Empty priority queue and hash set}
    \State $seq_{best} \gets candidate$
    \State $e_{best} \gets \Call{energy}{candidate}$
    \myspace
    \State $\mathbb{PQ}.\Call{push}{candidate}$
    \myspace 
    \State $u \gets 0$
    \myspace
    \While{$u < T_u$}
        \State $u \gets u + 1$
        \State $seq_{current} \gets \mathbb{PQ}.\Call{pop}$
        \myspace
        \For{$\textbf{each}\ i \in 0..L$} \Comment{Neighborhood search}
            \State $hash \gets \Call{hash\_flip}{i}$
            \medskip
            \If {$\mathbb{PQ}.\Call{contains}{hash}$}
                \State \textbf{continue}
            \EndIf
            \myspace
            \State $seq_{current} \gets \Call{sequence\_flip}{seq_{current}, i}$
            \myspace
            \State $\mathbb{PQ}.\Call{push}{seq_{current}}$
            \State $\mathbb{H}.\Call{insert}{hash}$
            \State $e_{current} \gets \Call{energy}{seq_{current}}$
            \myspace
            \If {$e_{current} < e_{best}$}
                \State $seq_{best} \gets seq_{current}$ \Comment{New best sequence found}
                \State $e_{best} \gets e_{current}$
                \State $u \gets 0$ \Comment{Reset $u$ to support dynamic depth}
            \EndIf
            \State $\Call{make\_rotations}{seq_{current}, \mathbb{PQ}, \mathbb{H}, T_r}$
            \State $seq_{current} \gets \Call{sequence\_flip}{seq_{current}, i}$
        \EndFor
    \EndWhile
    \myspace
    \State \Return $seq_{best}$ \Comment{Best sequence found}
\end{algorithmic}
\end{small}
\end{algorithm}


Until now, the search has exclusively involved sequences of odd lengths, a constraint dictated by the skew-symmetry rule, which is only defined for odd-length sequences. However, sequence operators can be deployed to extend the search to sequences of even lengths, as shown in~\cite{dimitrov22}, by appending or removing elements on each end of the sequence. The application of these operators yields new sequences of different lengths (odd or even) with comparable $F$ to the original odd-length sequences. These newly derived sequences were reintroduced as candidates to the second step, with the expectation of further lowering the energy $E$, as shown in Fig.~\ref{fig:flow}.

\begin{figure}
\centering
\includegraphics[width=0.7\linewidth]{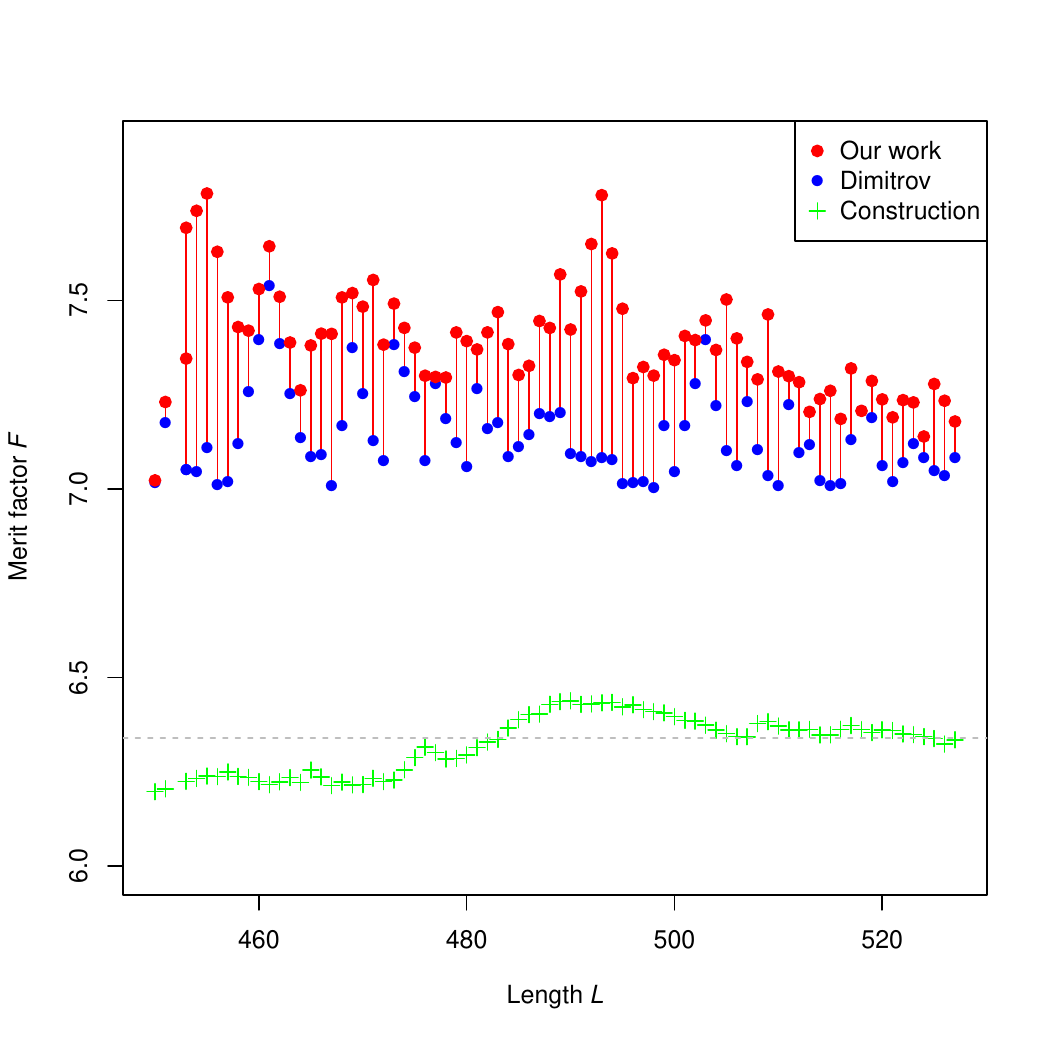}
\caption{\label{fig:record}Merit factor values ($F$) for binary sequences of length $450 \le L \le 527$. The dual-step approach found sequences with significantly higher merit factors compared to the previous best known values reported by Dimitrov~\cite{dimitrov22}. The merit factors of constructed sequences are also plotted for comparison.}
\end{figure}

\section{RESULTS}\label{sec:results}
We implemented the Algorithm~\ref{algo:saw} using the C++ programming language, CUDA Toolkit 12.2.0, and Clang 16.0.6 compiler. We employed the Vega\footnote{Vega is Slovenia’s petascale supercomputer -- \url{https://doc.vega.izum.si/}} grid computing environment powered by the NVIDIA A100-SXM4-40GB GPUs to evaluate its performance. Similarly, we implemented the Algorithm~\ref{algo:pq} in the Rust programming language utilizing the RustC 1.78.0 compiler, which ran on a 4-core ARM Neoverse-N1 device with 24 GB of memory running Ubuntu 24.04. Using this setup, we improved all results of the skew-symmetric sequences in just a few hours. Given that skew-symmetric sequences are odd-length, we used sequence operators to extend the improvements to include even-length sequences and then reintroducing them to the second step. This approach allowed for a comprehensive enhancement across both odd- and even-length sequences. 

The improvements we achieved over previously best reported merit factors, as detailed in~\cite{dimitrov22}, are depicted in Fig.~\ref{fig:record} and collected in Table~\ref{tab:meritFactors}. The graph illustrates the increase in the merit factor for sequences of $450 \le L \le 527$, with an average increase of $0.2543$ and a maximum observed increase of $0.6980$ for a sequence of length $493$, which reached a merit factor of $7.7791$. The highest merit factor was found for $L=455$ reaching $7.7835$. For comparative purposes, the graph also displays the merit factors of constructed sequences, further highlighting the effectiveness of our approach. Previous best-known merit factor values~\cite{dimitrov22} and new binary sequences are presented in Table~\ref{tab:meritFactors} by using hexadecimal coding. Each hexadecimal digit is presented with a binary string: $0 = 0000$, $1 = 0001$, ..., $F = 1111$. Therefore, it is necessary to remove the leading $0$ values to obtain the correct length of the binary sequence. To get the the sequence values, each $0$ of the binary string should be converted to $-1$.

To demonstrate the effectiveness of Algorithm~\ref{algo:pq}, we ran two experiments on an arbitrarily selected sequence of odd-length multiple times with the same limited number of self-avoiding walks and parameter values for $T_i$ and $T_u$ for the second experiment. The first experiment used only Algorithm~\ref{algo:saw}, and we recorded the  lowest energy for each run. The second experiment used both algorithms with sieving. To assess whether there was a statistically significant difference between the results obtained from the two experiments, we applied the Wilcoxon Rank Sum test to the $30$ independent observations of each experiment, selecting a significance level of $\alpha=0.05$. 
\begin{figure}[thb]
\centering
\includegraphics[width=0.6\linewidth]{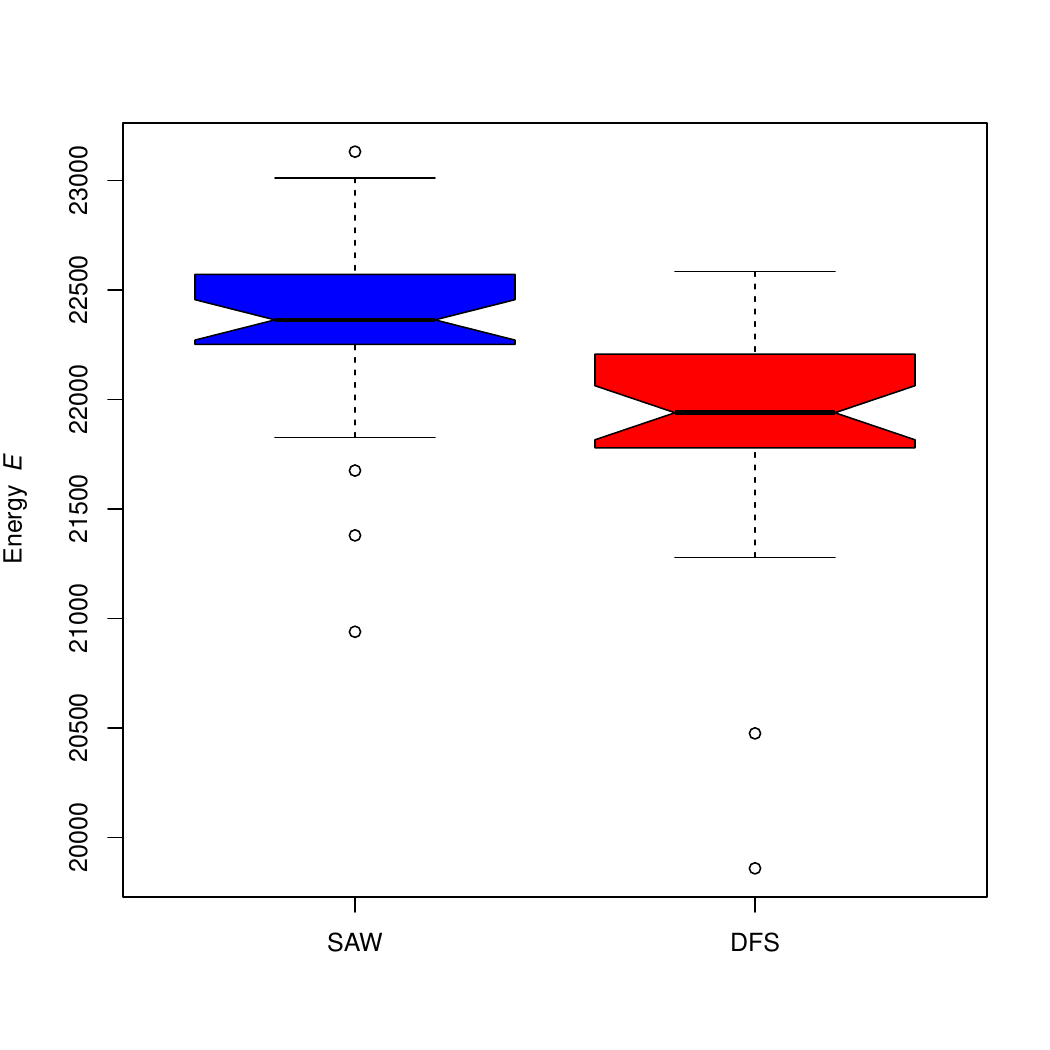}
\caption{\label{fig:boxplot}Box-plots of the experiments for $L=519$. The blue box-plot (SAW) shows the distribution of the best achieved energy values of each run using only the first step. The red box-plot (DFS) shows the distribution for the experiment using both steps which has a significantly lower median energy.}
\end{figure}
This non-parametric test was chosen due to the non-normal distribution of data, making it an appropriate choice for analyzing differences in central tendency between the two sets of experimental outcomes. The results of the Wilcoxon Rank Sum test revealed a highly significant difference between the two experimental conditions, with a p-value of less then $0.0041$, strongly indicating that the inclusion of Algorithm~\ref{algo:pq} leads to a notable improvement in the optimization process. Furthermore, the distribution of the energy values and the significant difference between the two experimental results are visually represented in the box-plot depicted in Fig.~\ref{fig:boxplot}, specifically for the sequence length $L=519$. It must be noted, that the notches of the two medians on the box-plots do not overlap, further indicating a significant difference. The box-plot clearly illustrates the lower energy achieved when both algorithms were employed in conjunction. This visual evidence, coupled with the statistical analysis, supports the conclusion, that Algorithm~\ref{algo:pq} contributes substantially to lowering the energy value, thus producing a higher merit factor, making it a valuable addition to the search process.

\subsection{Seeding the second step with Legendre sequences}
 As previously discussed, construction methods can efficiently generate sequences of arbitrary length with a merit factor of around $6.3421$ or $6.4382$. This is achieved by rotating a Legendre sequence of prime length by approximately one-quarter of its length and then appending the initial elements of the sequence to itself, as described in references~\cite{borwein04, Jedwab13}. We employed this methodology to construct sequences of a specified length by systematically varying both the number of appended elements ($0.055 \le \hat{t} \le 0.063$) and the rotation parameter ($0.20 \le r \le 0.24$) of the Legendre sequence. These values for parameters $\hat{t}$ and $r$ were selected based on the theoretical analysis in~\cite{borwein04}. The sequences generated through this process were subsequently used as candidates for Algorithm~\ref{algo:pq}. It is important to note that the sequences produced by this method are not skew-symmetric, so using them as seeds (random starting sequences) in the first step is not possible. Table~\ref{tab:construct} shows the best merit factor $F$ of a constructed sequence of length $L$ and the best merit factor found after seeding the second step (DFS) with Legendre sequences. 
\begin{table}[tbh]
\caption{The merit factor (\textit{F}) of the best Legendre sequence and the best merit factor found by DFS after seeding it with Legendre sequences of length $L$.}
\label{tab:construct}
\centering
\begin{tabular}{|c|c|c|}
    \hline
    \multirow{2}{*}{\textbf{Length} \textit{L}} & \multicolumn{2}{c|}{\textbf{Merit factor} \textit{F}} \\ 
    \cline{2-3}
    & \textbf{Construction} & \textbf{DFS} \\
    \hline
    450 & 6.1976 & 6.3088 \\        
    \hline
    471 & 6.2332 & 6.2998 \\        
    \hline
    491 & 6.4292 & 6.4292 \\        
    \hline
    526 & 6.3240 & 6.3240 \\     
    \hline
\end{tabular}
\end{table}
 It must be noted, that only some randomly selected lengths $L$ are tabulated. Despite the initial potential of these constructed sequences to yield relatively high merit factors, the optimization process resulted in minimal to no improvement. 
 This lack of enhancement in the merit factor is likely attributable to the fact that the sequences generated by the construction method reside in or near a local minimum in the energy landscape of the search space. The optimization algorithm struggles to escape this local minimum, which hinders its ability to effectively search for the global minimum. Consequently, the search process failed to produce sequences with significantly improved merit factors, despite the promising starting point offered by the constructed sequences.

\section{CONCLUSION}\label{sec:conclusions}
The problem of finding aperiodic low auto-correlation binary sequences (LABS) is a significant computational challenge. In this paper, we introduced a hybrid approach that effectively tackles this problem through two distinct steps. In the first step, we leverage skew-symmetry, restriction classes, and parallel computing on graphic processing units (GPUs) to efficiently explore the limited search space. The second step employs a priority queue to guide the search, enabling a comprehensive exploration of the entire problem space. This dual-step strategy has proven to be a powerful method for discovering sequences with higher merit factors.

Our approach led to the discovery of new best-known binary sequences for all lengths between 450 and 527, with one exception. For L = 518, the merit factor was matched with a different sequence. This demonstrates the effectiveness of our method in pushing the boundaries of known solutions.

Moreover, the experiment involving the seeding of the second step with Legendre sequences provided valuable insights. Although these sequences showed initial promise, they ultimately failed to produce significant improvements in the merit factor due to the presence of local minimum in the local search space. This result highlights the inherent limitations of purely constructive methods when used alongside optimization algorithms, underscoring the complexity of the LABS problem.

\section*{ACKNOWLEDGMENT}
The authors (J. B. and B. B.) acknowledge the financial support from the Slovenian Research and Innovation Agency (Research Core Funding No.~P2-0041 -- Computer Systems, Methodologies, and Intelligent Services). 
The authors would also like to acknowledge the Slovenian
Initiative for National Grid (SLING) for using computational resources for performing some experiments in this paper.

\balance
\bibliographystyle{unsrt}

\begin{thebibliography}{10}

\bibitem{golay72}
M.~Golay.
\newblock A class of finite binary sequences with alternate auto-correlation values equal to zero (corresp.).
\newblock {\em IEEE Transactions on Information Theory}, 18(3):449--450, 1972.

\bibitem{littlewood66}
J.~E. Littlewood.
\newblock On polynomials $\sum^n \pm z^m, \sum^ne^{a_mi}z^m,z=e^{0i}$.
\newblock {\em Journal of the London Mathematical Society}, s1-41(1):367--376, 1966.

\bibitem{jedwab2004survey}
Jonathan Jedwab.
\newblock A survey of the merit factor problem for binary sequences.
\newblock In {\em International Conference on Sequences and Their Applications}, pages 30--55. Springer, 2004.

\bibitem{katz2024moments}
Daniel~J Katz and Miriam~E Ramirez.
\newblock Moments of autocorrelation demerit factors of binary sequences.
\newblock {\em Designs, Codes and Cryptography}, pages 1--45, 2024.

\bibitem{Lib-OPUS-labs-1987-JourPhys-Bernasconi}
J.~Bernasconi.
\newblock Low autocorrelation binary sequences: statistical mechanics and configuration space analysis.
\newblock {\em J. Physsique}, 48:559--567, April 1987.

\bibitem{Mullen13HANDBOOKMERITFACTOR}
Gary~L. Mullen and Daniel Panario.
\newblock Other correlation measures.
\newblock In {\em Handbook of Finite Fields}, chapter 10.3.5, pages 322--324. Chapman \& Hall/CRC, 1st edition, 2013.

\bibitem{dimitrov22}
Miroslav Dimitrov.
\newblock New classes of binary sequences with high merit factor.
\newblock {\em arXiv preprint arXiv:2206.12070}, 2022.

\bibitem{borwein04}
P.~Borwein, K.-K.S. Choi, and J.~Jedwab.
\newblock Binary sequences with merit factor greater than 6.34.
\newblock {\em IEEE Transactions on Information Theory}, 50(12):3234--3249, 2004.

\bibitem{Jedwab13}
Jonathan Jedwab, Daniel~J. Katz, and Kai-Uwe Schmidt.
\newblock Advances in the merit factor problem for binary sequences.
\newblock {\em Journal of Combinatorial Theory, Series A}, 120(4):882--906, 2013.

\bibitem{Schmidt2016}
Kai-Uwe Schmidt.
\newblock Sequences with small correlation.
\newblock {\em Designs, Codes and Cryptography}, 78, 01 2016.

\bibitem{Mow2015}
Wai~Ho Mow, Ke-Lin Du, and Wei~Hsiang Wu.
\newblock New evolutionary search for long low autocorrelation binary sequences.
\newblock {\em IEEE Transactions on Aerospace and Electronic Systems}, 51(1):290--303, 2015.

\bibitem{Chen2022}
Yutao Chen and Ronghao Lin.
\newblock Computationally efficient long binary sequence designs with low autocorrelation sidelobes.
\newblock {\em IEEE Transactions on Aerospace and Electronic Systems}, 58(3):1966--1980, 2022.

\bibitem{Kimura2024}
Keigo Kimura and Hiroshi Takase.
\newblock A search method for binary codes compressed to several sub-pulses using quantum annealing and search experiment by ising machine.
\newblock {\em Journal of Signal Processing}, 28(4):103--106, 2024.

\bibitem{Packebusch16}
Tom Packebusch and Stephan Mertens.
\newblock Low autocorrelation binary sequences.
\newblock {\em Journal of Physics A: Mathematical and Theoretical}, 49(16), 2016.

\bibitem{Boskovic17}
Borko Bošković, Franc Brglez, and Janez Brest.
\newblock {Low-autocorrelation binary sequences: On improved merit factors and runtime predictions to achieve them}.
\newblock {\em Applied Soft Computing}, 56:262--285, 2017.

\bibitem{Boskovic24}
Borko Bošković, Jana Herzog, and Janez Brest.
\newblock Parallel self-avoiding walks for a low-autocorrelation binary sequences problem.
\newblock {\em Journal of Computational Science}, 77:102260, 2024.

\bibitem{Baden11}
John~Michael Baden.
\newblock {Efficient Optimization of the Merit Factor of Long Binary Sequences}.
\newblock {\em IEEE Transactions on Information Theory}, 57(12):8084--8094, 2011.

\bibitem{shaydulin2024evidence}
Ruslan Shaydulin, Changhao Li, Shouvanik Chakrabarti, Matthew DeCross, Dylan Herman, Niraj Kumar, Jeffrey Larson, Danylo Lykov, Pierre Minssen, Yue Sun, et~al.
\newblock Evidence of scaling advantage for the quantum approximate optimization algorithm on a classically intractable problem.
\newblock {\em Science Advances}, 10(22):eadm6761, 2024.

\bibitem{Farnane18}
Kaoutar Farnane, Khalid Minaoui, and Driss Aboutajdine.
\newblock Local search algorithm for low autocorrelation binary sequences.
\newblock In {\em 2018 4th International Conference on Optimization and Applications (ICOA)}, pages 1--5, 2018.

\bibitem{dimitrov21}
Miroslav Dimitrov.
\newblock {On the Skew-Symmetric Binary Sequences and the Merit Factor Problem}.
\newblock {\em ArXiv}, abs/2106.03377, 2021.

\bibitem{Halim08}
Steven Halim, Roland H.~C. Yap, and Felix Halim.
\newblock {Engineering Stochastic Local Search for the Low Autocorrelation Binary Sequence Problem}.
\newblock In Peter~J. Stuckey, editor, {\em Principles and Practice of Constraint Programming}, pages 640--645, Berlin, Heidelberg, 2008. Springer Berlin Heidelberg.

\bibitem{Gallardo09}
José~E. Gallardo, Carlos Cotta, and Antonio~J. Fernández.
\newblock Finding low autocorrelation binary sequences with memetic algorithms.
\newblock {\em Applied Soft Computing}, 9(4):1252--1262, 2009.

\bibitem{Brest18}
Janez Brest and Borko Bošković.
\newblock {A Heuristic Algorithm for a Low Autocorrelation Binary Sequence Problem With Odd Length and High Merit Factor}.
\newblock {\em IEEE Access}, 6:4127--4134, 2018.

\bibitem{Brest22}
Janez Brest and Borko Bošković.
\newblock {Computational Search of Long Skew-symmetric Binary Sequences with High Merit Factors}.
\newblock {\em MENDEL}, 28(2):17--24, Dec. 2022.

\bibitem{bloom70}
Burton~H. Bloom.
\newblock Space/time trade-offs in hash coding with allowable errors.
\newblock {\em Commun. ACM}, 13(7):422–426, jul 1970.

\bibitem{dimitrov20}
Miroslav Dimitrov, Tsonka Baitcheva, and Nikolay Nikolov.
\newblock {On the Generation of Long Binary Sequences With Record-Breaking PSL Values}.
\newblock {\em IEEE Signal Processing Letters}, 27:1904--1908, 2020.

\end{thebibliography}


\onecolumn



\end{small}

\twocolumn

\end{document}